\documentclass[12pt]{article}
\usepackage{amsfonts,amssymb,amsmath}
\usepackage[dvips]{epsfig}
\begin{document}
\title{{\em Approach of the Generating Functions to the Coherent States for Some Quantum Solvable Models}}
\author{B. Mojaveri\thanks{Email: bmojaveri@azaruniv.ac.ir}\hspace{2mm} and \hspace{2mm} A. Dehghani\thanks{Email: a\_dehghani@tabrizu.ac.ir,  alireza.dehghani@gmail.com}\\
{\small {\em Department of Physics, Azarbaijan Shahid Madani
University, PO Box 51745-406, Tabriz, Iran\,}}\\
{\small {\em Department of Physics, Payame Noor University, P.O.Box
19395-3697 Tehran, I.R. of Iran. \,}}} \maketitle
\begin{abstract}
We introduce to this paper new kinds of coherent states for some
quantum solvable models: a free particle on a sphere,
one-dimensional Calogero-Sutherland model, the motion of spinless
electrons subjected to a perpendicular magnetic field $B$,
respectively , in two dimensional flat surface and an infinite flat
band. We explain how these states come directly from the generating
functions of the certain families of classical orthogonal
polynomials without the complexity of the algebraic approaches. We
have shown that some examples become consistent with the Klauder-
Perelomove and the Barut-Girardello coherent states. It can be
extended to the non-classical, q-orthogonal and the exceptional
orthogonal polynomials, too. Especially for physical systems that
they don't have a specific algebraic structure or involved with the
shape invariance
symmetries, too.\\\\
{\bf PACS Nos:}  02.30.Gp, 02.20.Sv, 42.50.Dv, 03.65-w, 05.30. - d,
03.65.Fd, 03.65.Ge

\end{abstract}

\section{Reviews and Motivation}
According to the pioneering work of Schr\"{o}dinger
\cite{Schrodinger} coherent states in its general form can be
realized as Gaussian wave-function could be constructed from a
particular superposition of the wave functions corresponding to the
discrete eigenvalues of the harmonic oscillator. They play an
important role in quantum optics and provide us with a link between
quantum and classical oscillators. Here, it is necessary to
emphasize that quantum coherence of states nowadays pervades many
branches of physics such as quantum electrodynamics, solid-state
physics, and nuclear and atomic physics, from both theoretical and
experimental viewpoints. Successfully , these states were applied to
some other models considering with their Lie algebra symmetries by
Glauber \cite{Glauber}, Klauder \cite{Klauder}, Sudarshan
\cite{Sudarshan}, Barut and Girardello \cite{Barut} and Perelomov
\cite{Perelomov}. Additionally, for the models with one degree of
freedom either discrete or continuous spectra- with no remark on the
existence of a Lie algebra symmetry- Gazeau et al proposed new
coherent states, which was parametrized by two real parameters
\cite{Gazeau}. Also, there exist some considerations in connection
with coherent states corresponding to the shape invariance
symmetries \cite{Fukui}.

To construct coherent states, four main different approaches— the
so-called Schr\"{o}dinger, Klauder-Perelomov( K-P),
Barut-Girardello( B-G), and Gazeau-Klauder( G-K) methods— have been
found, so the second and the third approaches rely directly on the
Lie algebra symmetries and their corresponding generators. Where the
first of them are established only by means of generating function-
regardless of the Lie algebraic symmetries.

It is worth to mention that the reverse idea is well known. In other
words coherent states may usually be led to the generating functions
attached to the classical polynomials, too \cite{Spiridonov}. Using
generating function of the Hermite polynomials $H_{n}(x)$
\cite{Grad}
\begin{eqnarray}
&&\hspace{-14mm}G(x,t)=e^{2xt-t^2}=\sum^{\infty}_{n=0}{\frac{t^n}{n!}H_{n}(x)},\hspace{4mm}|t|<\infty
\end{eqnarray}
then with complexification of $t\sqrt{2}\longrightarrow z\in
\mathbb{C}$, one gets
\begin{eqnarray}
&&\hspace{-24mm}G(x,z)=e^{xz\sqrt{2}-\frac{z^2}{2}}=\pi^{1/4}e^{\frac{x^2}{2}}\sum^{\infty}_{n=0}{\frac{z^n}{\sqrt{n!}}\hspace{2mm}\phi_{n}(x)}\nonumber\\
&&\hspace{-10mm}=\pi^{1/4}e^{\frac{x^2+|z|^2}{2}}\hspace{2mm}\langle
x\mid z\rangle_{Sch},\end{eqnarray}where we have used the notations
$\phi_{n}(x) \left(
=\frac{e^{-\frac{x^{2}}{2}}}{\sqrt{2^{n}n!\sqrt{\pi}}}\hspace{1mm}H_{n}(x)\right)$
as the complete and orthonormal eigenvectors of the simple harmonic
oscillator( SHO), i.e.
\begin{eqnarray}
&&\hspace{-24mm}\int^{\infty}_{-\infty}{\phi_{n}(x)\phi_{m}(x)dx}=\delta_{nm},\nonumber\end{eqnarray}
and $\langle x\mid z\rangle_{Sch} \left(
=e^{-\frac{|z|^2}{2}}\sum^{\infty}_{n=0}{\frac{z^n}{\sqrt{n!}}\hspace{1mm}\phi_{n}(x)}\right)$
as the Schr\"{o}dinger type of coherent states\footnote{ Subscript
`Sch' refers to the Schr\"{o}dinger and specifies a particular type
of quantum states that are called the canonical coherent states,
too.} attached to it, respectively. Clearly, Eq. (2) indicates that
using generating function of the classical orthogonal polynomials
one can easily found\footnote{However, as will be discussed later,
this can not be applied to all of the classical orthogonal
polynomials.} the coherency of quantum states, i.e.
\begin{eqnarray} &&\hspace{-24mm}\langle x\mid
z\rangle_{Sch}=\frac{e^{-\frac{x^2+|z|^2}{2}}}{\pi^{1/4}}\hspace{2mm}G(x,z).\end{eqnarray}
It, also, serves the following identities
\begin{eqnarray} &&\hspace{-26mm}\hat{a}\langle x\mid
z\rangle_{Sch}=z\langle x\mid
z\rangle_{Sch},\\
&&\hspace{-24mm}\langle x\mid
z\rangle_{Sch}=e^{z\hat{a}^{\dag}-\overline{z}\hat{a}}\phi_{0}(x),\end{eqnarray}
with respect to the boson creation and annihilation operators
$\hat{a}^{\dag}$ and $\hat{a}$. They represent commutators of the
Heisenberg Lie algebra, i.e. $[\hat{a}, \hat{a}^{\dag}] = 1$, on the
Hilbert space of the orthonormal eigenvectors of the harmonic
oscillator, $\mathcal{H} := span\{\phi_{n}(x):
\int^{\infty}_{-\infty}{dx\phi_{n}(x)\phi_{n'}(x)} =
\delta_{nn'}\}\mid^{\infty}_{n=0}$, and realize the usual laddering
relations $\hat{a}^{\dag}\phi_{n} = \sqrt{n+1}\phi_{n+1}$ and
$\hat{a}\phi_{n}
= \sqrt{n}\phi_{n-1}$.\\
{{\em In fact, Eq (3) together with the Eqs. (4) and (5) refer to
the fact that the above mentioned approaches the Schr\"{o}dinger,
B-G and K-P come to the same results when applied to the simple
harmonic oscillator. It should be noted that these methods have
different and distinct results for other physical models,
respectively}}.

Following the extension of the above idea of other classical
orthogonal polynomials, we have tried it on several cases including
{ the associated Legendre polynomials $P^{m}_{l}(x)$ as well as the
associated Laguerre polynomials $L^{m}_{n}(x)$ and the
associated Bessel functions $B^{q, \beta}_{n, m}$} \cite{Fakhri5}.\\
Because of the relations between the associated Legendre functions
$P^{m}_{l}(x)$ and the spherical harmonics $Y_{lm}\left(\theta,
\phi\right)$, they correspond to the quantum states of many physical
problems and have, also, been studied in the framework of the
coherent states theory (see, for example, Refs. [12- 24]).
Therefore, our motivation in this paper is to make new kind of
coherent states as infinite superpositions of the spherical
harmonics which are based only on the structure of the corresponding
generating function of no remark on the existence of the Lie algebra
symmetries. Here, we want to emphasize that the people could not
think about constructing of this type of coherent states. The
reverse idea is previously discussed and realization of the
Klauder-Perelomov types of coherent states as finite superpositions
of the spherical harmonics associated to the unitary irreducible
representations of the compact algebra $su(2)$ via orbital angular
momentum operators is years old, also realization of the
Barut-Girardello as well as the Klauder-Perelomov types of coherent
states as infinite superpositions of the spherical harmonics
corresponding to the infinite-dimensional representations of the Lie
algebra $su(1,1)$ \cite{Fakhri4}, have recently been introduced
\cite{Fakhri1, Fakhri2}. Also, Fakhri. et. al established new type
of the coherent states in accordance with the new generating
functions associated with the spherical harmonics \cite{Fakhri3}. It
should be noted that different generating functions attached to the
associated Legendre polynomials of different quantum numbers $l$ and
$m$, for example \cite{Arfken}
\begin{eqnarray}
&&\hspace{-14mm}G_{m}(x,t)=\frac{(2m)!(1-x^2)^{m/2}}{2^mm!(1+t^2-2xt)^{m+1/2}}=\sum^{\infty}_{l=0}{{t^l}P^{m}_{l+m}(x)},\hspace{4mm}|t|<1.\end{eqnarray}
Along with the application of the second type of the generating
functions (6), we construct new kind of coherent states as infinite
superpositions of the spherical harmonics in the section (2). Which
is distinct from the the Barut-Girardello as well as the
Klauder-Perelomov types of coherent states discussed earlier in Refs
\cite{Fakhri1, Fakhri2, Fakhri3}.

Another part of this document, that will become in section (3), is
spent to design and analysis of the coherent states are derived from
the generating functions of the associated Laguerre polynomials
$L^{m}_{n}$ \cite{Grad}
\begin{eqnarray}
&&\hspace{-14mm}G^{+}_{m}(x,z)=(xz)^{-\frac{m}{2}}J_{m}(2\sqrt{xz})e^{z}=\sum^{\infty}_{n=0}{\frac{z^n}{(n+m)!}L^{m}_{n}(x)},\hspace{6mm}m>-1\\
&&\hspace{-14mm}G^{-}_{m}(x,z)=\frac{e^{-\frac{xz}{1-z}}}{(1-z)^{m+1}}=\sum^{\infty}_{n=0}{{z^n}L^{m}_{n}(x)},\hspace{6mm}|z|<1,\\
&&\hspace{-14mm}G^{0}_{m}(x,z)=(1+z)^{m}e^{-xz}=\sum^{\infty}_{n=0}{{z^n}L^{m-n}_{n}(x)},\hspace{6mm}|z|<1.\end{eqnarray}
Because of there exist some quantum physical models whose
solvability and square integrability are connected with the
associated Laguerre polynomials. The Landau levels, Morse potential,
Calogero-Sutherland model, half- oscillator, radial part of 3-D
harmonic oscillator and radial part of a hydrogen-like atom are the
six examples [25- 30]. They play an important role in the quantum
mechanics and have been considered by many authors in the framework
of the coherent states theory \cite{Kahn, Deh1, Deh2}. We will show
three different kinds of the generating functions (7), (8) and (9)
lead to the well known Barut-Girardello and the Klauder-Perelomov
types of coherent states respectively, already obtained in Refs
\cite{Deh1, Deh2} due to the one dimensional Calogero-Sutherland
model as well as two dimensional Landau levels. In other words the
generating function technique presented here, come to the same
results with the B-G and the K-P methods when applied to the Landau
Levels and the Calogero-Sutherland model. It should be noted that in
the generating function formalism we don't require an algebraic
structure and don't deal with the complexity of algebraic methods.

The last part is devoted to the Bessel functions
$B^{(q,\beta)}_{l,m} (x)$. The square-integrable associated Bessel
functions can be applied to obtain bound states corresponding to
some one-dimensional super-symmetric potential and also some
two-dimensional quantum-mechanical models having a Lie algebra
symmetry of $su(1,1)$ \cite{Fakhri4, Fakhri5, Fakhri6}. Therefore,
exponential generating functions corresponding to the formal power
series of associate Bessel functions $B^{(q,\beta)}_{l,m} (x)$ of
the same $l$, the same $l + m$ and the same $l- m$, are important
not only from the point of view of mathematical derivation but also
from the point of view of physical applications.
\section{Coherent states attached to the associated Legendre
polynomials $P^{m}_{l}(x)$} Let us remind that the spherical
harmonics in terms of the polar and azimuthal angles, $0 \leq \theta
\leq \pi$ and $0 \leq \varphi \leq 2\pi$, can be written as:
\begin{eqnarray}
&&\hspace{-14mm}\langle\theta, \varphi\mid l,m\rangle:=
Y^{m}_{l}(\theta,
\varphi)=(-1)^m\sqrt{\frac{2l+1}{4\pi}\frac{(l-m)!}{(l+m)!}}e^{im\varphi}P^{m}_{l}(\cos\theta),\hspace{2mm}l\in
\mathbb{N}_{0},\hspace{2mm}-l\leq m \leq l.\end{eqnarray} They
constitute an infinite dimensional Hilbert space $\mathcal{H}=
L^{2}\left(S^{2}, d\Omega\left(\theta, \phi\right)\right)$,
\begin{eqnarray}
&&\hspace{-14mm}\langle l, m|l', m'\rangle:=
\int_{S^2}{{Y^{m}_{l}}^{\ast}(\theta, \varphi)Y^{m'}_{l'}(\theta,
\varphi)d\Omega\left(\theta,
\phi\right)}=\delta_{ll'}\delta_{mm'}.\end{eqnarray} The bases of
$\mathcal{H}$ are independent spherical harmonics on the sphere
$S^2$ with different values for both indices $l$ and $m$. The
properties of the spherical harmonics are well known and may be
found in many texts and papers (see, for example, Refs. [23- 29]).
The spherical harmonics with a given $l$ are wave functions of a
free particle which has to be placed on the surface of the sphere of
a given energy \cite{Chenaghlou}. With the application of angular
momentum operators, a given unitary irreducible representation of
$so(3)\cong su(2)$ Lie algebra is characterized as the spherical
harmonics with a given $l$. This is the most well-known property of
them. It is worth to mention, for decades the spherical harmonics,
$Y_{lm}\left(\theta, \phi\right)$ have been considered as the
representation space of compact Lie algebra $su(2)$ and recently
extended by Fakhri to noncompact Lie algebra $su(1,1)$
\cite{Fakhri4}.

Motivate by this situation, in this paper we will consider the
general generating functions $G(\theta, t)$ of the associated
Legendre functions $P^{m}_{l}$, are given by \cite{Arfken}
\begin{eqnarray}
&&\hspace{-14mm}G(\theta,
t)=\frac{(2m)!}{{2^m}m!}\frac{(\sin\theta)^m}{(1+t^2-2t\cos\theta)^{m+1/2}}\nonumber\\
&&\hspace{-1mm}=\sum^{\infty}_{l=0}{t^{l}P^{m}_{m+l}(\cos\theta)},\hspace{2mm}|t|\leq
1.\end{eqnarray} The above expression after restoring the Eq. (11),
provide us to evaluate furthered sums involving spherical harmonics
\begin{eqnarray}
&&\hspace{-14mm}\frac{(-1)^me^{im\varphi}}{\sqrt{4\pi}}G(\theta,
t)=\sum^{\infty}_{l=0}{{t^{l}}\sqrt{\frac{(l+2m)!}{l!(2l+2m+1)}}Y^{m}_{m+l}(\theta,
\varphi)}.\end{eqnarray} Along with substitution of
$t\longrightarrow \mathfrak{z}\in \mathbb{C}$, it becomes
\begin{eqnarray}
&&\hspace{-14mm}\frac{(-1)^me^{im\varphi}}{\sqrt{4\pi}}G(\theta,
\mathfrak{z})=\langle\theta,
\varphi\mid\left(\sum^{\infty}_{l=0}{{\mathfrak{z}^{l}}\sqrt{\frac{(l+2m)!}{l!(2l+2m+1)}}\mid
m+l, m\rangle}\right).\end{eqnarray} We shall show that the
expression of right hand sides of the Eq. (16) has all the features
of coherent states. For this reason we introduce new kind of
coherent states of spherical harmonics as follows
\begin{eqnarray}
&&\hspace{-14mm}|\mathfrak{z}\rangle_{m}=M_{m}^{-1/2}(|\mathfrak{z}|)
\sum^{\infty}_{l=m}{{\mathfrak{z}^{l}}\sqrt{\frac{(l+m)!}{(l-m)!(2l+1)}}\mid
l, m\rangle},\end{eqnarray} where $M^{-1/2}_{m}(|\mathfrak{z}|)$ is
chosen so that $|\mathfrak{z}\rangle_{m}$ is normalized, i.e.
$_{m}\langle \mathfrak{z} \mid \mathfrak{z} \rangle_{m}= 1$. Due to
the orthogonality relation (13) follow that overlapping of two
different kinds of these normalized states become nonorthogonal, if
$\mathfrak{z}'\neq \mathfrak{z}$, i.e.
\begin{eqnarray}
&&\hspace{-14mm}_{m}\langle \mathfrak{z'} \mid \mathfrak{z}
\rangle_{m}=\frac{(3m)!}{(2m+1)}\frac{\left(\overline{\mathfrak{z'}}\mathfrak{z}\right)^m}
{\sqrt{M_{m}(|\mathfrak{z'}|)M_{m}(|\mathfrak{z}|)}}\hspace{2mm}{_2F_1}\left(\left[3m+1,
m+\frac{1}{2}\right],
\left[m+\frac{3}{2}\right],\overline{\mathfrak{z'}}\mathfrak{z}\right)
.\end{eqnarray} Then, $M_{m}(|\mathfrak{z}|)$ can be calculated to
be taken as
\begin{eqnarray}
&&\hspace{-14mm}M_{m}(|\mathfrak{z}|)=\frac{(3m)!}{(2m+1)}|\mathfrak{z}|^{2m}\hspace{2mm}{_2F_1}\left(\left[3m+1,
m+\frac{1}{2}\right],
\left[m+\frac{3}{2}\right],|\mathfrak{z}|^{2}\right).\end{eqnarray}
From the completeness relation of the Fock sub-space states,
resolution of the identity condition
\begin{eqnarray}&&\hspace{-5mm}\oint_{\mathbb{C}(\mathfrak{z})}{|{\mathfrak{z}}\rangle_{m}
\hspace{1mm}{_{m}\langle
{\mathfrak{z}}|}}d\mu_{m}(|\mathfrak{z}|)=I_{m}=\sum_{l=m}^{\infty}{|l,m\rangle\langle
l, m|},\end{eqnarray} is realized for the states
$|{\mathfrak{z}}\rangle_{m}$ with $d\mu_{m}(|\mathfrak{z}|) :=
\mathfrak{K}_{m}(|\mathfrak{z}|)
\frac{d|\mathfrak{z}|^{2}}{2}d\phi$, satisfied with the following
positive definite and non-oscillating measures on a unit disc:
\begin{eqnarray}&&\hspace{-14mm}\mathfrak{K}_{m}(|\mathfrak{z}|)=
\frac{M_{m}(|\mathfrak{z}|)}{\pi(2m-2)!}(1+r^{-2})(1-r^{-2})^{2m-2}\nonumber\\
&&\hspace{-11mm}=\frac{(3m)!}{\pi(2m+1)(2m-2)!}\frac{(1+r^2)(1-r^{2})^{2m}{_2F_1}\left(\left[3m+1,
m+\frac{1}{2}\right],
\left[m+\frac{3}{2}\right],r^{2}\right)}{(1-r^{2})^{2}r^{2m-2}},\end{eqnarray}
where we have used the polar coordinate of $\mathfrak{z}=re^{i\phi},
\hspace{2mm}0\leq r\leq1,\hspace{2mm}0\leq \phi\leq2\pi$. We want to
emphasize, the coherent states $|\mathfrak{z}\rangle_{m}$ don't
obtained through of the B-G nor the K-P methods. We naturally expect
that these states will result some new quantum and statistical
features.
\section{Coherent states attached to the associated
Laguerre polynomials $L^{n}_{m}(x)$} In this section, we construct
some coherent states according to the one dimensional
Calogero-Sutherland model and the two dimensional Landau Levels are
emerged through of the three different kinds of the generating
functions of the associated Laguerre polynomials Eqs. (7), (8) and
(9).\\\\
$\bullet$ {\textbf{Coherent states arising from the first generating function $G^{+}_{m}(x, z)$ in Eq. (7)}}:\\
Here, we review the solvability of the Calegero-Sutherland
Hamiltonian $H^{\lambda}$ on the half-line $x$
\begin{eqnarray}
&&\hspace{-1.5cm}H^{\lambda}=\frac{1}{2}\left[-\frac{d^2}{dx^2}+x^2+\frac{\lambda(\lambda-1)}{x^2}\right],\end{eqnarray}
where, the simple an-harmonic term $\frac{\lambda(\lambda-1)}{x^2}$
refers to the Goldman-Krivchenkov potential \cite{Fu}. In Refs.
\cite{ Deh1, Hall}, it has been shown that the second-order
differential operators
\begin{eqnarray}
&&\hspace{-1.5cm} {J}_{\pm}^{\lambda}:=\frac{1}{4}\left[\left(x\mp\frac{d}{dx}\right)^2-\frac{\lambda(\lambda-1)}{x^2}\right],\\
&&\hspace{-1.5cm}{J}_{3}^{\lambda}:=\frac{H^{\lambda}}{2},\end{eqnarray}
satisfy the standard commutation relations of $su(1,1)$ Lie algebra
as follows
\begin{eqnarray}
&&\hspace{-1.5cm}\left[{J}_{+}^{\lambda},{J}_{-}^{\lambda}\right]=-2{J}_{3}^{\lambda},
\hspace{20mm}\left[{J}_{3}^{\lambda},{J}_{\pm}^{\lambda}\right]=\pm{J}_{\pm}^{\lambda}.
\end{eqnarray}
Also, product the unitary and positive-integer irreps of $su(1,1)$
Lie algebra as
\begin{eqnarray}&&\hspace{-15mm}{J}_{+}^{\lambda}|n-1,\lambda\rangle=\sqrt{n\left(n+\lambda-\frac{1}{2}\right)}|n,\lambda\rangle,\\
&&\hspace{-15mm}{J}_{-}^{\lambda}|n,\lambda\rangle=\sqrt{n\left(n+\lambda-\frac{1}{2}\right)}|n-1,\lambda\rangle,\\
&&\hspace{-15mm}{J}_{3}^{\lambda}|n,\lambda\rangle=\left(n+\frac{\lambda}{2}+\frac{1}{4}\right)|n,\lambda\rangle.\end{eqnarray}
We assume that the set of states described above, form complete and
orthonormal basis
of an infinite dimensional Hilbert space, i.e. \begin{eqnarray}&&\hspace{-15mm}\mathcal{H}^{\lambda}:={\mathrm{span}}\{|n,\lambda\rangle|\langle n, \lambda|m, \lambda\rangle=\delta_{nm}\}|_{n=0}^{\infty},\nonumber\\
&&\hspace{-15mm}\langle
x|n,\lambda\rangle:=(-1)^n\sqrt{\frac{2\Gamma(n+1)}{\Gamma(n+\lambda+\frac{1}{2})}}x^{\lambda}
e^{-\frac{x^2}{2}}L_{n}^{\lambda-\frac{1}{2}}(x^2),
\hspace{4mm}\lambda >\frac{-1}{2},\end{eqnarray} where
$L_{n}^{\lambda-\frac{1}{2}}(x)$ denotes the associated Laguerre
polynomials \cite{Grad}. Along with the orthogonality of the
associated Laguerre polynomials, the orthogonality relation of the
basis of $\mathcal{H}^{\lambda}$ reads
\begin{eqnarray}&&\hspace{-15mm}\langle n, \lambda|m, \lambda\rangle:=\frac{2n!}{\Gamma(n+\lambda+\frac{1}{2})}
\int_{0}^{\infty}x^{2\lambda}e^{-x^2}L_{n}^{\lambda-\frac{1}{2}}(x^2)
L_{m}^{\lambda-\frac{1}{2}}(x^2)dx=\delta_{nm}.\end{eqnarray}

Using Eqs. (7) and (27), one gets
\begin{eqnarray}
&&\hspace{-44mm}\sqrt{2}x^{\lambda}
e^{-\frac{x^2}{2}}G^{+}_{\lambda-\frac{1}{2}}(x^2,z)=\sqrt{2}x^{\frac{1}{4}}z^{\frac{1-2\lambda}{4}}
e^{z-\frac{x^2}{2}}J_{\lambda-\frac{1}{2}}(2x\sqrt{z})\nonumber\\
&&\hspace{-4mm}=\sum^{\infty}_{n=0}{\frac{(-z)^n}{\sqrt{n!\Gamma(n+\lambda+\frac{1}{2})}}\langle
x|n,\lambda\rangle},\hspace{6mm}\lambda>-\frac{1}{2}.\end{eqnarray}
Obviously, R.H.S of (29) is proportional to the Barut-Giradello
coherent states for the Calegero-Sutherland model already discused
in \cite{Deh1} ( see Eq. (6) therein) i.e.
\begin{eqnarray}
&&\hspace{-44mm}\sum^{\infty}_{n=0}{\frac{(-z)^n}{\sqrt{n!\Gamma(n+\lambda+\frac{1}{2})}}\langle
x|n,\lambda\rangle}\equiv\langle x\mid
-z\rangle^{\lambda}_{BG},\end{eqnarray} and satisfies an eigenvalue
equation with respect to the lowering operator ${J}_{-}^{\lambda}$.
{\em We conclude that the generating function of the associated
Laguerre polynomials (7) is proportional to the Barut-Giradello
coherent states for the Calegero-Sutherland model up to a
normalization coefficient}, i.e.
\begin{eqnarray}
&&\hspace{-44mm}\langle x\mid
-z\rangle^{\lambda}_{BG}\equiv\sqrt{2}x^{\lambda}
e^{-\frac{x^2}{2}}G^{+}_{\lambda-\frac{1}{2}}(x^2,z).\end{eqnarray}It
indicates that the generating function $G^{+}_{m}(x, z)$ in Eq. (7),
come to the B-G coherent states when applied to the Calogero-Sutherland model.\\\\
$\bullet$ {\textbf{Coherent states arising from the second generating function $G^{-}_{m}(x, z)$, Eq. (8)}}:\\
Similarity to what we have performed above, one can show that
\begin{eqnarray} &&\hspace{-44mm}\sqrt{2}x^{\lambda}
e^{-\frac{x^2}{2}}G^{-}_{\lambda-\frac{1}{2}}(x^2,z)=\sqrt{2}\frac{x^{\lambda}
e^{-\frac{x^2}{2}(1+2\frac{z}{1-z})}}{(1-z)^{\lambda+\frac{1}{2}}}\nonumber\\
&&\hspace{-4mm}=\sum^{\infty}_{n=0}{{(-z)^n}{\sqrt{\frac{\Gamma(n+\lambda+\frac{1}{2})}{n!}}}\langle
x|n,\lambda\rangle}\\
&&\hspace{-4mm}\equiv\langle x\mid
-z\rangle^{\lambda}_{KP}.\end{eqnarray} Clearly, it illustrates a
correspondence between the generating function of the associated
Laguerre polynomials (8) and the Klauder-Perelomve coherent states
for the Calegero-Sutherland model( Eq. (11) in Ref. [29]).\\\\
$\bullet$ {\textbf{Coherent states arising from the third generating function $G^{0}_{m}(x, z)$ in Eq. (9)}}:\\
Likewise, it is easy to see that the third generating function
$G^{0}_{m}(x, z)$
\begin{eqnarray} &&\hspace{-30mm}G^{0}_{m}(x,z)=(1+z)^{m}e^{-xz}
=\sum^{\infty}_{n=0}{{z^n}L^{m-n}_{n}(x)}
=z^{m}\sum^{\infty}_{n=-m}{{z^n}L^{-n}_{n+m}(x)},\hspace{6mm}|z|<1,\end{eqnarray}
with respect to the polar coordinate, $0 < r <\infty, 0\leq \varphi<
2\pi$, representation of Landau levels in terms of the associated
Laguerre functions
\begin{eqnarray}
&&\hspace{-1.5cm}\langle r, \varphi|n+m,
-n\rangle=\sqrt{\frac{(n+m)!\left(\frac{M\omega}{2\hbar}\right)^{-n+1}}{\pi
n!}}
r^{-n}e^{-in\varphi}e^{-\frac{M\omega}{4\hbar}r^2}L^{(-n)}_{n+m}\left(\frac{{M\omega
r^2}}{{2\hbar}}\right),\end{eqnarray}  becomes
\begin{eqnarray} &&\hspace{-40mm}\sqrt{\frac{1}{2\pi}}z^{-m}e^{-{M\omega
r^2}/{4\hbar}}G_{m}\left({M\omega
r^2}/{2\hbar},z\right)=\left(1+\frac{1}{z}\right)^{m}e^{-\frac{M\omega
r^2}{4\hbar}(1+2z)}\nonumber\\
&&\hspace{-10mm}=\sum^{\infty}_{n=-m}{{\left(ze^{i\phi}\sqrt{{\frac{M\omega
r^2}{2\hbar}}}\right)^n}\sqrt{\frac{m!}{(m+n)!}}\langle r,
\varphi|n+m, -n\rangle}.\end{eqnarray}Here, the Landau levels are
related to the symmetric-gauge Landau Hamiltonian\footnote{The
Landau Hamiltonian, $H$, has an infinite-fold degeneracy on the
Landau levels, that is in which Landau cyclotron frequency is
expressed in terms of the value of the electron charge, its mass,
the magnetic field strength $B_{ext}$ and also the velocity of light
as $\omega= \frac{eB_{ext}}{Mc}$. Here $m$ is an integer and $n$ a
nonnegative integer, subject to the constraint $m\geq -n$.}
corresponding to the motion of an electron on the flat surface in
the presence of an unified magnetic field, $B_{ext}$, in the
positive direction of $z$ axis \cite{Kahn, Deh2}, i.e.
\begin{eqnarray}
&&\hspace{-1.5cm} H |n,m\rangle= \hbar\omega\left(n +
\frac{1}{2}\right) |n,m\rangle,\\
&&\hspace{-1.5cm}
H=\hbar\omega\left({a}^{\dag}a+\frac{1}{2}\right)=\hbar\omega\left({b}^{\dag}b+\frac{1}{2}\right)-\omega
L_{3},\end{eqnarray} where $L_{3} =
-i\frac{\partial}{\partial\varphi}$ and
\begin{eqnarray}&&\hspace{-1.5cm}a= -e^{i\varphi}
\sqrt{\frac{\hbar}{2M\omega}}\left(\frac{\partial}{\partial
r}+\frac{i}{r}\frac{\partial}{\partial
\varphi}+\frac{M\omega}{2\hbar}r\right),\hspace{4mm} {a}^{\dag}=
e^{-i\varphi}
\sqrt{\frac{\hbar}{2M\omega}}\left(\frac{\partial}{\partial
r}-\frac{i}{r}\frac{\partial}{\partial
\varphi}-\frac{M\omega}{2\hbar}r\right),\\
&&\hspace{-1.5cm}b= e^{-i\varphi}
\sqrt{\frac{\hbar}{2M\omega}}\left(\frac{\partial}{\partial
r}-\frac{i}{r}\frac{\partial}{\partial
\varphi}+\frac{M\omega}{2\hbar}r\right),\hspace{4mm} {b}^{\dag}=
-e^{i\varphi}
\sqrt{\frac{\hbar}{2M\omega}}\left(\frac{\partial}{\partial
r}+\frac{i}{r}\frac{\partial}{\partial
\varphi}-\frac{M\omega}{2\hbar}r\right).\end{eqnarray}They form two
separate copies of Weyl-Heisenberg algebra,
\begin{eqnarray}
&&\hspace{-1.5cm}[a, {a}^{\dag}] = 1, [b, b^{\dag}] = 1, [a,
b^{\dag}] = [a^{\dag}, b] = [a, b] = [a^{\dag}, b^{\dag}]
=0,\end{eqnarray} with the unitary representations as
\begin{eqnarray}
&&\hspace{-1.5cm} a |n,m\rangle = \sqrt{n} |n - 1,m + 1\rangle
,\hspace{4mm}
a^{\dag} |n - 1,m + 1\rangle = \sqrt{n} |n,m\rangle ,\\
&&\hspace{-1.5cm} b |n,m\rangle = \sqrt{n + m} |n,m - 1\rangle
,\hspace{4mm} b^{\dag} |n,m -1\rangle = \sqrt{n + m}
|n,m\rangle.\end{eqnarray} Also, they are complex conjugate of each
other with respect to the following  orthogonality integration over
the entire plane,
\begin{eqnarray}
&&\hspace{-1.5cm}\langle n, m|n',
m'\rangle:=\int^{2\pi}_{\varphi=0}\int^{\infty}_{r=0}{\langle r,
\varphi|n, m\rangle^{\ast}\langle r, \varphi|n', m'\rangle
rdrd\varphi}=\delta_{nn'}\delta_{mm'}.\end{eqnarray}

Applying the annihilation operator $a$ on {\em the R.H.S} of the Eq.
(36), then using the laddering relation (42) one can find that it
satisfies {\em an eigenvalue equation}, which implies a
proportionality among these coherent states emerged through of the
generating function, $G^{0}_{m}(x, z)$, of the associated Laguerre
polynomials and the normalized Weyl-Heisenberg coherent states,
$|ze^{i\phi}\sqrt{{\frac{M\omega r^2}{2\hbar}}}\rangle^{b}_{n}$,
i.e.
\begin{eqnarray} &&\hspace{-40mm}\sqrt{\frac{1}{2\pi}}z^{-m}e^{-{M\omega
r^2}/{4\hbar}}G_{m}\left({M\omega r^2}/{2\hbar},z\right)\equiv
\langle r,\varphi\mid ze^{i\phi}\sqrt{{\frac{M\omega
r^2}{2\hbar}}}\rangle^{b}_{n}.\end{eqnarray}Relevant characteristics
such as resolution of the identity condition as well as their
statistical properties were investigated in Refs. \cite{Kahn, Deh2}.
\section{Coherent states attached to the associated
Bessel functions $B^{\mu, \nu}_{n, m}(x)$} \textbf{The Model}:\\\\
Consider a spinless electron with charge $e < 0$ and an effective
mass $\mu$ moving on an infinite flat band of the presence of a
Morse-like perpendicular magnetic fields directed in the negative
$z$-direction $B = -B_{0}e^{-\frac{2\pi}{ao}x}\hat{k}$ with $B_{0} >
0$ as a constant magnetic strength and $a_{0}$ as the width of the
band. It leads to the following Morse-like vector potential
\cite{Fakhri6}
\begin{eqnarray} &&\hspace{-40mm}A_{x}=\frac{i\pi\hbar c}{ea_{0}}-i\frac{a_{0}B_{0}}{2\pi}e^{-\frac{2\pi}{a_{0}}x},
\hspace{4mm}A_{y}=\frac{a_{0}B_{0}}{2\pi}e^{-\frac{2\pi}{a_{0}}x},\hspace{4mm}A_{z}=0,\end{eqnarray}
in which $c$ is the velocity of electromagnetic waves in the vacuum.
For an electron of effective mass $\mu$, moving on the infinite flat
band in the presence of the vector potential (46), the
time-independent Schr\"{o}dinger wave equation
\begin{eqnarray} &&\hspace{-40mm}\frac{1}{2\mu}\left[\left(p_{x}-\frac{e}{c}A_{x}\right)^2
+\left(p_{y}-\frac{e}{c}A_{y}\right)^2\right]\psi=E\psi,\nonumber\end{eqnarray}
can be written in terms of the variables
$\xi=e^{\frac{2\pi}{a_{0}}x}$ ($0 < \xi < \infty$) and
$-\frac{a_{0}}{2} < y < \frac{a_{0}}{2}$ as
\begin{eqnarray} &&\hspace{-30mm}\left[-{\xi^{2}}\frac{\partial^2}{\partial\xi^2}+
\left(\frac{eB_{0}a_{0}^2}{2\pi^2\hbar
c}-2\xi\right)\frac{\partial}{\partial\xi}-\frac{a_{0}^2}{4\pi^2}\frac{\partial^2}{\partial
y^2}+i\frac{eB_{0}a_{0}^3}{4\pi^3\hbar c}\frac{\partial}{\xi\partial
y}-\frac{1}{4}\right]\psi=\frac{2\mu
a_{0}^2}{4\pi^2\hbar^2}E\psi.\end{eqnarray} The periodic boundary
condition in the $y$-direction requires that the wave function
$\psi$ is separated into $\psi = e^{\frac{2i\pi}{a_{0}}my}
\psi(\xi)$. Hence, $\psi(\xi)$ satisfies the following differential
equation:
\begin{eqnarray} &&\hspace{-20mm}{\xi^{2}}\frac{d^2\psi(\xi)}{d\xi^2}+
\left(2\xi-\frac{eB_{0}a_{0}^2}{2\pi^2\hbar
c}\right)\frac{d\psi(\xi)}{d\xi}-\left(m^2-\frac{1}{4}
+\frac{eB_{0}a_{0}^2}{2\pi^2\hbar c}\frac{m}{\xi}-\frac{2\mu
a_{0}^2}{4\pi^2\hbar^2}E\right)\psi(\xi)=0.\end{eqnarray} Which can
be compared by the associated Bessel differential equation, and
results \begin{eqnarray} &&\hspace{-20mm}\psi(\xi)=B^{0, \beta}_{l,
m}(\xi).\end{eqnarray} Here, $B^{(q,\beta)}_{l, n} (x)$ refer to the
Bessel functions, with $\beta=-\frac{eB_{0}a_{0}^2}{2\pi^2\hbar c}$.
Therefore, the square integrable solutions can be obtained as follow
\begin{eqnarray} &&\hspace{-20mm}|l, m\rangle:= \psi_{l, m}(x,
y)=\frac{\beta}{\sqrt{2\pi}} e^{\frac{2i\pi}{a_{0}}my}B^{0,
\beta}_{l, m}(e^{\frac{2\pi}{a_{0}}x}),\end{eqnarray} They form an
orthonormal set with respect to the integer index $m$,
\begin{eqnarray} &&\hspace{-20mm}\langle l, m|l', m'\rangle:=
\int^{\frac{a_{0}}{2}}_{-\frac{a_{0}}{2}}{\int^{\infty}_{-\infty}{\psi^{\ast}_{l,
m}(x, y)\psi_{l', m'}(x, y)e^{\frac{2\pi}{a_{0}}x-\beta
e^{-\frac{2\pi}{a_{0}}x}}}dxdy}=\delta_{ll'}\delta_{mm'}.\end{eqnarray}
We also find that the allowed energies of the electron are quantized
as a positive quadratic function of both quantum numbers $l$ and
$m$:
\begin{eqnarray} &&\hspace{-20mm}E_{l, m}=
\frac{2\pi^2\hbar^2}{\mu
a_{0}^2}\left(m-l-\frac{1}{2}\right)\left(m+l+\frac{1}{2}\right).\end{eqnarray}
Therefore, the energy values increase in decreasing width $a_{0}$,
without increasing the strength of magnetic fields. In addition,
according to our considerations, there is no degeneracy in the case
where $(2m - 2l - 1) (2m + 2l + 1)$ is a prime number and two folds
in other cases. Also, the linear spectrum of the Landau levels is
obtained as a limiting case of $a_{0}\rightarrow \infty$.\\It has
been shown that the square integrable pure states realize
representations of $su(1, 1)$ algebra via the quantum number $n$
corresponding to the linear momentum in the $y$-direction. All of
the lowest states of the $su(1, 1)$ representations minimize
uncertainty relation and the minimizing of their second and third
states is transformed to that of the Landau levels in the limit
$a_{0}\rightarrow \infty$. The compact forms of the Barut-Girardello
coherent states corresponding to Irreducible representation of
$su(1, 1)$ algebra and their positive definite measures on the
complex plane are also calculated \cite{Fakhri6}.

What we do, here, is to consider the generating functions for given
$l + m$ and $q=0$, the case the sequences are increasing from $l$.
the sequences are increasing with respect to $l$. Due to whether
$n:=-l-m-1$ is odd or even, i.e. $n = 2k + 1$ or $n = 2k$, the
highest functions are $B^{(q,\beta)}_ {-k-1, \frac{q}{2}-k-1}(x)$
and $B^{(0,\beta)}_ {-k, -k-1}(x)$, respectively. These functions
lie on the lines $m = l$ and $m = l -1$ of figure 1 in
\cite{Fakhri5}, respectively. Therefore, it is obvious that the
terminology of highest functions has been devoted to the associated
Bessel functions $B^{(0, \beta)}_{ l,m} (x)$ with the most value of
$m$.

$\bullet$ We suppose that $n$ is even, i.e. $n = 2k$. For a given
value of $k$, the generating functions corresponding to the
second-type series will result an appropriate infinite sequence of
the associated Bessel functions
\begin{eqnarray}
&&\hspace{-40mm}G_{n=2k}(x, t)=\frac{x^{-k}}{2\sqrt{xt}}
\left[\left(1-\sqrt{xt}\right)^{2k}e^{\beta\sqrt{\frac{t}{x}}}
-\left(1+\sqrt{xt}\right)^{2k}e^{-\beta\sqrt{\frac{t}{x}}}\right]\nonumber\\
&&\hspace{-20mm}=\sum^{\infty}_{m=0}{\frac{t^m}{(2m+1)!}\frac{B^{(0,\beta)}_
{m-k, -m-k-1}(x)}{a_{m-k, -m-k-1}(0, \beta)}},\end{eqnarray} with
\[ a_{l, m}(0,
\beta) = \left\lbrace
  \begin{array}{c l}
    \frac{(-1)^{-m}\beta^{-l}}{\sqrt{(l-m)!(-l-m-1)!}} & \text{if $m\leq
l <0$},\\
    \frac{(-1)^{-l-m-1}\beta^{-l-1}}{\sqrt{(l-m)!(-l-m-1)!}} & \text{if $0\leq l
\leq -m-1$}.
  \end{array}
\right. \] It should be noticed that $G_{n=2k}(x)$ is summed over
the parameter $m$ for $l+m=-2k-1$.\\Thereafter, using $x$-coordinate
representation of the Bessel functions as well as the relation (50),
it yields
\begin{eqnarray}
&&\hspace{-20mm}\frac{\beta
e^{\frac{-2i\pi}{a_{0}}(k+1)y}}{\sqrt{2\pi(2k)!}}G_{2k}(x,
t)=\langle x\mid\left(\sum^{\infty}_{m=0}{{\frac{\left(\beta t
e^{\frac{2i\pi}{a_{0}}y}\right)^m}{\sqrt{(2m+1)!}}}}\mid m-k,
-m-k-1\rangle\right).\end{eqnarray} Where the Fock states $\mid m-k,
-m-k-1\rangle$ will require the following completeness and square
integrability conditions, respectively, as follows:
\begin{eqnarray} &&\hspace{-10mm}\sum^{\infty}_{m=0}\mid m-k,
-m-k-1\rangle\langle m'-k, -m'-k-1\mid = I_{2k},\\
&&\hspace{-10mm}\langle m-k, -m-k-1|m'-k, -m'-k-1\rangle:=\nonumber\\
&&\hspace{15mm}\int^{\pi}_{-\pi}{\int^{\infty}_{-\infty}{\psi^{\ast}_{m-k,
-m-k-1}(x, y)\psi_{m'-k, -m'-k-1}(x, y)e^{-\beta
e^{-\frac{2\pi}{a_{0}}x}}}dxdy}=\delta_{mm'}.\end{eqnarray} Here,
$I_{2k}$ refers to the identity( projection) operators on the
separable infinite dimensional Hilbert sub-spaces
$\mathcal{H}_{2k}:=span\{\mid m-k, -m-k-1\rangle\}$. One can show
that the orthogonality relation (56) can be readily obtained with
the aid of the generating function (54). This provides us the
corresponding CSs to be of the form
\begin{eqnarray}
&&\hspace{-20mm}\mid \mathfrak{z}\rangle_{2k}\left( = \mid \beta t
e^{\frac{2i\pi}{a_{0}}y}\rangle\right):=\sqrt{\frac{|\mathfrak{z}|}{\sinh(|\mathfrak{z}|)}}
\sum^{\infty}_{m=0}{{\frac{\mathfrak{z}^m}{\sqrt{(2m+1)!}}}}\mid
m-k, -m-k-1\rangle\end{eqnarray} ,and can be regarded as an infinite
series of solutions to the Hamiltonian of a spinless electron moving
on an infinite flat band of the presence of a Morse-like
perpendicular magnetic fields. Here, $\mathfrak{z}=\beta t
e^{\frac{2i\pi}{a_{0}}y}$ as an arbitrary complex variable and
coefficients $\sqrt{\frac{|\mathfrak{z}|}{\sinh(|\mathfrak{z}|)}}$
are included to achieve the normalization condition. It is
straightforward that, such a superposition contains all the required
features of a coherent state. In other word, one can show that these
normalized states form an 'overcomplete' system in the separable
Hilbert sub-spaces $\mathcal{H}_{2k}$. Indeed, they solve the
identity,
\begin{eqnarray}&&\hspace{-5mm}\oint_{\mathbb{C}(\mathfrak{z})}{|{\mathfrak{z}}\rangle_{2k}
\hspace{1mm}{_{2k}\langle
{\mathfrak{z}}|}}d\mu_{k}(|\mathfrak{z}|)=I_{2k}\nonumber\end{eqnarray}
in terms of an acceptable measure $d\mu(|\mathfrak{z}|) :=
\mathfrak{K}(|\mathfrak{z}|) \frac{d|\mathfrak{z}|^{2}}{2}d\phi$ on
the whole complex plane, with
\begin{eqnarray}&&\hspace{-17mm}\mathfrak{K}(|\mathfrak{z}|)=
\frac{e^{-\mathfrak{z}}}{2\pi|\mathfrak{z}|}\sinh(|\mathfrak{z}|).\end{eqnarray}
$\bullet$ Similar results can be found on the generating function
$G_{n=2k+1}(x, t)$ which is summed over the parameter $m$ for
$l+m=-2k-2$ and relates to the second-type sequences as an infinite
sequences of the associated Bessel functions
\begin{eqnarray}
&&\hspace{-40mm}G_{n=2k+1}(x, t)=\frac{x^{-k-1}}{2}
\left[\left(1-\sqrt{xt}\right)^{2k+1}e^{\beta\sqrt{\frac{t}{x}}}
+\left(1+\sqrt{xt}\right)^{2k+1}e^{-\beta\sqrt{\frac{t}{x}}}\right]\nonumber\\
&&\hspace{-20mm}=\sum^{\infty}_{m=0}{\frac{t^m}{(2m)!}\frac{B^{(0,\beta)}_
{m-k-1, -m-k-1}(x)}{a_{m-k-1, -m-k-1}(0, \beta)}}.\end{eqnarray} One
can show that this leads to the following normalized coherent vector
\begin{eqnarray}
&&\hspace{-20mm}\mid
\mathfrak{z}\rangle_{2k+1}=\frac{1}{\sqrt{\cosh(|\mathfrak{z}|)}}
\sum^{\infty}_{m=0}{{\frac{\mathfrak{z}^m}{\sqrt{(2m)!}}}}\mid
m-k-1, -m-k-1\rangle,\end{eqnarray} where we have used the
orthogonality relation
\begin{eqnarray} &&\hspace{-15mm}\langle m-k-1, -m-k-1\mid m'-k-1, -m'-k-1\rangle:=\nonumber\\
&&\hspace{10mm}\int^{\pi}_{-\pi}{\int^{\infty}_{-\infty}{\psi^{\ast}_{m-k-1,
-m-k-1}(x, y)\psi_{m'-k-1, -m'-k-1}(x, y)e^{-\beta
e^{-\frac{2\pi}{a_{0}}x}}}dxdy}=\delta_{mm'}.\end{eqnarray} Using
the completeness relation $\sum^{\infty}_{m=0}\mid m-k-1,
-m-k-1\rangle\langle m'-k-1, -m'-k-1\mid= I_{2k+1}$, it is found
that the relation
\begin{eqnarray}&&\hspace{-5mm}\oint_{\mathbb{C}(\mathfrak{z})}{|{\mathfrak{z}}\rangle_{2k+1}
\hspace{1mm}{_{2k+1}\langle
{\mathfrak{z}}|}}d\mu_{k}(|\mathfrak{z}|)=I_{2k+1}\nonumber\end{eqnarray}is
satisfied, through the following positive definite measure\\
\begin{eqnarray}&&\hspace{-5mm}d\mu_{k}(|\mathfrak{z}|)=
\frac{e^{-\mathfrak{z}}}{2\pi}\cosh(|\mathfrak{z}|)\frac{d|\mathfrak{z}|^{2}}{2}d\phi\hspace{5mm}|\mathfrak{z}|\in[0,
\infty), \hspace{4mm}0\leq\phi\leq 2\pi.\end{eqnarray}
\section{Discussion and Outlooks} We construct new and generalized coherent
states associated to the one and two dimensional quantum solvable
models and study some of their mathematical properties. We discuss
an algorithm that leads to superposition of quantum states play an
important role in quantum physics as coherent states. Despite of
well know techniques that are based on the group and representation
theory, this is only introduces approaches will focus on generating
functions. It results new and different types of coherent states
corresponding to the Legendre polynomials and the Bessel functions
too, which is discussed for first time then expects to explore some
unrevealed features. Our formalism allows us to calculate new kind
of coherent states especially for physical systems that they don't
have a specific algebraic structure or involved with the shape
invariance symmetries, too. Finally, it can be applied to some
non-classical and q-deformed polynomials of other quantum systems
and will be reported in a future work.

\end{document}